\documentclass[12pt,preprint]{aastex}
\usepackage{amsmath}
\usepackage{graphicx}
\usepackage{color}
\usepackage{url}
\usepackage{CJKutf8}
\usepackage{wasysym}
\usepackage{ulem}

\slugcomment{Accepted by the Astronomical Journal}

\shorttitle{Non-Gravitational Active Asteroids}
\shortauthors{Hui \& Jewitt}

\begin{document}

\title{Non-Gravitational Acceleration of the Active Asteroids}
\author{
Man-To Hui 
\begin{CJK}{UTF8}{bsmi}
(許文韜)$^{1}$ 
\end{CJK}
and David Jewitt$^{1,2}$
}
\affil{$^1$Department of Earth, Planetary and Space Sciences,
UCLA, 
595 Charles Young Drive East, 
Los Angeles, CA 90095-1567\\
}
\affil{$^2$Department of Physics and Astronomy, UCLA, 
430 Portola Plaza, Box 951547, Los Angeles, CA 90095-1547\\
}
\email{pachacoti@ucla.edu}

\begin{abstract}

Comets can exhibit non-gravitational accelerations caused by recoil forces due to anisotropic mass loss. So might active asteroids. We present an astrometric investigation of 18 active asteroids in search of non-gravitational acceleration. Statistically significant (signal-to-noise ratio (SNR) $> 3$) detections are obtained in three objects: 313P/Gibbs, 324P/La Sagra and (3200) Phaethon. The strongest and most convincing detection ($>$7$\sigma$ in each of three orthogonal components of the acceleration), is for the $\sim$1 km diameter nucleus of 324P/La Sagra. A 4.5$\sigma$ detection of the transverse component of the acceleration of 313P/Gibbs (also $\sim$1 km in diameter) is likely genuine too, as evidenced by the stability of the solution to the rejection or inclusion of specific astrometric datasets. We also find a 3.4$\sigma$ radial-component detection for $\sim$5 km diameter (3200) Phaethon, but this detection is more sensitive to the inclusion of specific datasets, suggesting that it is likely spurious in origin. The other 15 active asteroids in our sample all show non-gravitational accelerations consistent with zero. We explore different physical mechanisms which may give rise to the observed non-gravitational effects, and estimate mass-loss rates from the non-gravitational accelerations. We present a revised momentum-transfer law based on a physically realistic sublimation model for future work on non-gravitational forces, but note that it has little effect on the derived orbital elements.
  
\end{abstract}


\section{\uppercase{Introduction}}

Active asteroids have the dynamical characteristics of asteroids but exhibit transient mass loss, resulting in the production of comet-like appearance (Hsieh and Jewitt 2006).  A working definition is that they are bodies which present evidence of mass loss, have semimajor axes, $a$, smaller than Jupiter's semimajor axis, and have Tisserand parameter with respect to Jupiter, $T_\mathrm{J} \ge 3.08$. There are currently $\sim$20 known active asteroids.  A number of mechanisms drive the mass loss, including the likely sublimation of exposed ice, asteroid-asteroid impact, and rotational disruption probably driven by radiation torques (Jewitt 2012; Jewitt et al.~2015).

The dynamics of active asteroids are of particular interest. Numerical simulations have been conducted to study the dynamical stability of some of these objects (c.f. Jewitt et al. 2015 and citations therein). Recent work by Hsieh \& Haghighipour (2016) investigated orbital evolution of test particles dynamically close to the $T_\mathrm{J} \simeq 3$ boundary between asteroids and comets. They found that, due to gravitational interactions with terrestrial planets and temporary trapping by mean-motion resonances with Jupiter, the fraction of the Jupiter-family comets fortuitously evolved into main-belt like orbits on Myr timescales could be as large as $\sim$0.1--1\%. However, most such main-belt captures would be transient, and long-term stable orbits with both small eccentricities and inclinations should be much more rare. 

Non-gravitational accelerations, if present, might significantly influence the dynamics of small bodies. Fern\'andez et al. (2002) and Levison et al.~(2006) found that capture into comet 2P/Encke's orbit is possible when assisted by plausible non-gravitational forces from outgassed material, but takes much longer than the expected outgassing lifetimes of comets. They suggested that 2P/Encke might have completed this capture while spending most of its time in a dormant state. Forces due to photon momentum (the Yarkovsky effect (e.g., Chesley et al. 2003; Vokrouhlick\'y et al. 2008; Chesley et al. 2012; Nugent et al. 2012; Farnocchia et al. 2014) and radiation pressure) are expected to be tiny compared to forces resulting from protracted anisotropic mass loss but have been detected in small asteroids.

To date, the only independently reported measurement of non-gravitational acceleration due to outgassing in an active asteroid is a 3$\sigma$ detection for 133P/(7968) Elst-Pizarro (Chesley et al. 2010a). In order to develop a better understanding of the active asteroids, we attempt to measure their non-gravitational accelerations. 

\section{\uppercase{Data Analysis and Method}}
\label{method}

Marsden et al. (1973) developed a standard orbit determination technique with non-gravitational effects. The non-gravitational acceleration of a small body, in terms of its radial (i.e., in the antisolar direction), transverse, and normal components $\mathcal{A_{\mathrm{R}}}$, $\mathcal{A}_{\mathrm{T}}$, and $\mathcal{A}_{\mathrm{N}}$, is related to three non-gravitational parameters $A_{j}$ ($j = 1, 2, 3$), which are expressed in the same right-handed Cartesian orthogonal coordinates system by

\begin{equation}
\left(
\begin{array}{c}
\mathcal{A}_{\mathrm{R}} \\
\mathcal{A}_{\mathrm{T}} \\
\mathcal{A}_{\mathrm{N}}
\end{array}
\right)
= \left(
\begin{array}{c}
A_{1} \\
A_{2} \\
A_{3}
\end{array}
\right) \cdot g\left(r \right),
\label{eq1}
\end{equation}

\noindent where $g\left( r \right)$ is the dimensionless standard momentum-transfer law at heliocentric distance, $r$, in AU. Marsden et al. (1973) defined $g(r)$ as:

\begin{equation}
g\left(r \right) = \alpha \left(\frac{r}{r_{0}} \right)^{-m} \left[ 1 + \left( \frac{r}{r_{0}} \right)^{n} \right]^{-k},
\label{eq2}
\end{equation}

\noindent in which $m = 2.15$, $n = 5.093$, $k = 4.6142$, the scaling distance $r_{0} = 2.808$ AU, and the normalisation factor $\alpha = 0.111262$, such that $g = 1$ at $r = 1$ AU. Accelerations $\mathcal{A}_j$ and $A_j$ are traditionally expressed in AU day$^{-2}$.  The momentum-transfer law comes from the assumption by Marsden et al. (1973) that the non-gravitational acceleration of a small body is proportional to the rate of sublimation of water-ice on an isothermal nucleus, with the momentum-transfer law reflecting the proportionality, such that the non-gravitational parameters $A_{j}$ are always constant. (Sublimation of other materials such as sodium and forsterite can be approximated by the same formalism with different parameters (c.f. Sekanina \& Kracht 2015), but the sublimation rates of these much less volatile materials are negligible compared to that of water.) In keeping with previous work, we proceed by assuming that the momentum-transfer law due to isothermal water-ice sublimation gives rise to the non-gravitational effects of the active asteroids.

We downloaded astrometric observations of all the active asteroids from the Minor Planet Center (MPC) Database Search\footnote{\url{http://www.minorplanetcenter.net/db_search}}, and then employed \textit{Find\_Orb} by B. Gray for orbit determination. The code uses numerical ephemeris DE431, and includes relativistic effects due to the gravity of the Sun, and perturbations by the eight major planets. Pluto and the thirty most massive  asteroids\footnote{The masses of the 30 most massive asteroids range from $\sim$$7 \times 10^{18}$ kg (375 Ursula) to $9 \times 10^{20}$ kg (1 Ceres). The values are based on the BC-405 asteroid ephemeris by Baer et al. (2011). } are also included. Astrometric observations were debiased and weighted as described in Farnocchia et al. (2014) and Chesley et al. (2010b) before orbit determination. 

We first calculated purely gravitational orbital solutions for each of the active asteroids, assuming $A_j = 0$ ($j = 1,2,3$). Weights would be relaxed to be comparable with corresponding ad hoc astrometric residuals. We next rejected astrometric observations whose residuals were greater than $\pm$3\arcsec.0 from ad hoc osculating solutions, in an iterative manner. For main-belt objects, such residuals are  large compared to systematic errors from the timing or  plate constant solutions. They may result from centroiding errors possibly due to the faintness or non-stellar appearance of the object,  from interference with background sources or adjacent cosmic rays or from other, unspecified errors. The threshold was chosen to exclude bad outliers while keeping as many data points as possible. Next, we included $A_{j}$ ($j=1,2,3$) as free parameters to be obtained from the best fit orbital solutions. The procedures for filtering outliers and relaxing weights were applied iteratively until convergence was achieved. This normally took three to five runs, somewhat dependent upon the quality of data. We finally recorded the converged orbital solutions along with $A_{j}$ $(j=1,2,3)$. 


\section{\uppercase{Results}}
\label{results}

We summarize the resulting non-gravitational parameters of the active asteroids in Table \ref{tab_ng}. Included are statistically confident detections (SNR $> 3$) of non-gravitational accelerations for 324P/La Sagra in all the three components, for (3200) Phaethon in the radial direction, and for 313P/Gibbs in the transverse direction. The other active asteroids show no statistically significant evidence (SNR $\le 3$) for non-gravitational effects. 

Our non-detection of the radial component of non-gravitational acceleration in 133P/(7968) Elst-Pizarro contradicts a 3$\sigma$ detection reported by Chesley et al. (2010a). However, if only observations prior to 2011 are considered, our result becomes similar to that of Chesley et al. (2010a). Therefore, we conclude that the reported detection is tied to the specific astrometric dataset employed, and cannot be trusted as real. Likewise, active asteroid 259P/Garradd shows marginal evidence of a radial non-gravitational acceleration with SNR = 2.97 (see Table \ref{tab_ng}). However, the result is found to change wildly depending on the particular astrometric observations selected. Moreover, the fit to 259P/Garradd relies on the smallest number of observations (40, compared to hundreds or thousands for other objects in Table \ref{tab_ng}). Therefore, we do not regard it as a significant detection.

\subsection{313P/Gibbs}
Hui \& Jewitt (2015)  previously  discussed the non-gravitational motion of this $\sim$1 km diameter object. We did not debias the astrometric observations and simply set equal weights to all the data. Nevertheless, the result is consistent with the one in the present work in which we employed more stringent techniques to weight the data. In this sense, the detection of $A_2$, at 4.5$\sigma$ confidence (Table \ref{tab_ng}) is relatively insensitive to the method by which the astrometric observations are handled. We thus conclude that it is likely a genuine detection of the transverse non-gravitational acceleration. Admittedly, in order to strengthen this conclusion, more observations of the object are desirable.

\subsection{324P/La Sagra}
324P/La Sagra shows the strongest non-gravitational acceleration of all the active asteroids, with detections $>$7$\sigma$ in all three components (see Table \ref{tab_ng}). The solutions are unlikely to be caused by contamination from undetected systematics in the astrometry because random exclusions of large subsets of the astrometric data hardly change the result. For example, discarding all the data from 2015 leads to no change in the significance of the $A_j$ parameters. Other tests, including arbitrary assignment of equal weights to all the data, have been made, without materially changing the result. While the detection of non-gravitational acceleration appears to be secure, the solution is nevertheless somewhat puzzling. In particular, the radial component, $A_1$, is negative (radial non-gravitational acceleration towards the Sun), which seems physically unrealistic in the context of sublimation from the hot day-side of the nucleus. This may indicate that the applied momentum-transfer law by Marsden et al. (1973) is inappropriate to this case, because the mass-loss rate does not vary symmetrically with heliocentric distance (or, equivalently, perihelion time) as described by Equation (\ref{eq2}) (see Figure 6 in Jewitt et al. (2016)). Another possibility is that it suggests a circumpolar or high-latitude active source and certain combinations of the spin-axis orientation of its nucleus (Yeomans et al. 2004).

\subsection{(3200) Phaethon}
Since the discovery in 1983, asteroid (3200) Phaethon had never been observed to show any signs of activity until 2009, 2012 and 2016 when it brightened by a factor of two around perihelion detected by the Solar Terrestrial Relations Observatory (STEREO) spacecraft (Jewitt \& Li 2010; Li \& Jewitt 2013; Hui \& Li 2016). Intriguingly, we have a $\mbox{SNR} = 3.4$ detection for its radial non-gravitational parameter $A_1$, which is statistically significant. Tests such as discarding all observations prior to 1990, or applying an equal weight scheme do affect the SNR slightly, but always leave $\mbox{SNR} \sim 3$. However, we can destroy the significance of the detection by, for instance, discarding all the data from the  discovery epoch to the mid-1990s. Alternatively, if a much stricter cutoff for astrometric residuals is employed (e.g. $\lesssim 1\arcsec.5$), resulting in removing observations overwhelmingly from the 1980s and early 1990s, the SNR shrinks to $\sim$2 and thus $A_1$ becomes insignificant. We therefore take the conservative position that the radial non-gravitational component is likely spurious. This is supported by the observation that (3200) Phaethon remains inactive until it is close to the Sun, where the activity is likely triggered by some process (thermal fracture, desiccation?) other than the sublimation of water ice (Jewitt \& Li 2010). 

\section{\uppercase{Discussion}}

\subsection{Test of the Procedure}

We conducted another test of the algorithms used by the orbit determination code $\textit{Find\_Orb}$ to be sure that the software does not introduce false detections of non-gravitational motion. For this purpose, we selected a dozen asteroids $\sim$10 km in diameter and having apparent magnitudes, orbits and observational histories similar to the majority of the active asteroids.  The 10 km asteroids, being $\sim$10$^3$ times more massive than the mostly $\sim$1 km scale active asteroids (Table \ref{tab_phys}), are unlikely to exhibit any measurable non-gravitational acceleration and thus serve as tests of the orbital fitting.  A list of candidates was generated by the JPL Small-Body Database Search Engine\footnote{\url{http://ssd.jpl.nasa.gov/sbdb_query.cgi}. Data retrieved on 2016 July 14.}. We applied the same procedures and techniques described in Section \ref{method} to obtain orbital solutions including $A_j$ ($j=1,2,3$) as free parameters. The results are summarized in Table \ref{tab_ast10km}.

As expected, none of the asteroids show significant  ($>$3$\sigma$) non-gravitational parameters. Some of the active asteroids have  fewer observations than have the selected moderate sized asteroids. We therefore truncated all the observations prior to 2010 for each of these asteroids and re-performed orbit determination. Again none shows detections on the non-gravitational parameters with SNR $> 3$. This confirms past work done with \textit{Find\_Orb} (e.g., Micheli et al.~2014) independently showing the reliability of the code. The validity of our cutoff set at SNR = 3 is justified as well. 

\subsection{Mass-Loss Estimates}

The mass-loss rate needed to provide a given non-gravitational acceleration can be estimated thanks to momentum conservation, using 
\begin{equation}
\dot{M} \left( t \right) = -\frac{M \left( t \right) g \left( r \left( t \right) \right) \sqrt{ A_{1}^{2} + A_{2}^{2} + A_{3}^{2}}}{\kappa \left( t \right) v \left( t \right)}
,
\label{eq_mloss}
\end{equation}

\noindent where  $M$ is the mass of the body, $v$ is the outflow speed of the ejecta, and $\kappa$ is a dimensionless factor which accounts for the collimation efficiency. The latter lies in the range $0 \le \kappa \le 1$, with $\kappa = 0$ for isotropic ejection and $\kappa = 1$ for perfectly collimated mass loss. We approximate the outflow speed as a function of heliocentric distance by mean thermal speed $v_{\mathrm{th}} = \sqrt{8 k_{\mathrm{B}} T / \left( \pi \mu m_{\mathrm{H}} \right)}$, where $\mu = 18$ is the molecular mass for the water-ice sublimation scenario, $m_{\mathrm{H}} = 1.67 \times 10^{-27}$ kg is the mass of the hydrogen atom and $k_{\mathrm{B}} = 1.38 \times 10^{-23}$ J K$^{-1}$ is the Boltzmann constant. We solve for the surface temperature, $T$, using the energy balance equation

\begin{equation}
\frac{\left(1 - \mathrm{A} \right) S_{\odot}}{r^{2}} \cos \zeta = \epsilon \sigma T^{4} + L\left( T \right) Z\left( T \right)
\label{eq_sub}
\end{equation}

\noindent in combination with the Clausius-Clapeyron relation for water ice. Here, $\mathrm{A}$ is the Bond albedo, $S_{\odot} = 1361$ W m$^{-2}$ is the solar constant, $\cos \zeta$ is the effective projection factor for the surface, $r$ is expressed in AU, $\epsilon$ is the emissivity, $\sigma = 5.67 \times 10^{-8}$ W m$^{-2}$ K$^{-4}$ is the Stefan-Boltzmann constant, $L\left( T \right)$ in J kg$^{-1}$ is the latent heat of vaporization, and $Z \left(T\right)$ in molecules per unit time per unit area is the gas production rate per unit area of surface. In this study, we assume $\epsilon = 1$, and $\cos \zeta = 1/4$, the latter corresponding to an isothermal nucleus, while $L (T)$ is documented in Huebner et al. (2006). The Bond albedos of the active asteroids are computed according to their geometric albedos by following the method by Bowell et al. (1989). The choice of $\cos \zeta = 1/4$ is made to remain consistent with the isothermal assumption by Marsden et al. (1973) (but see Appendix \ref{marsden}). 

The collimation efficiency remains observationally unconstrained, although observations showing that cometary emissions are largely sunward suggest that small values of $\kappa$ are unrealistic. We choose $\kappa \equiv 0.8$ for the sake of definiteness. Combined with Equation (\ref{eq_sub}), the time-average mass-loss rate around the orbit can be numerically estimated by transforming Equation (\ref{eq_mloss}) to

\begin{equation}
\overline{\dot{M}} \simeq -\frac{\pi \rho D^{3} \sqrt{ A_{1}^{2} + A_{2}^{2} + A_{3}^{2}}}{6\kappa P} \int_{0}^{P} \frac{g \left( r \left( t \right) \right)}{v_{\mathrm{th}} \left( r\left( t \right) \right)} \mathrm{d} t
,
\label{eq_mloss2}
\end{equation}

\noindent where $\rho$ is the bulk density, $D$ is the diameter of the body, and $P$ is the orbital period. We assume nominal density $\rho = 10^{3}$ kg m$^{-3}$  for all the active asteroids, while $D$ is extracted from either the JPL Small-Body Database Browser or Table 2 in Jewitt et al. (2015). The results are listed in Table \ref{tab_phys}. We calculated the uncertainty of $\overline{\dot{M}}$ solely from the covariance matrix of $A_j$ ($j=1,2,3$) based upon error propagation. For cases where objects have SNR $\le 3$ for $\overline{\dot{M}}$, we list 5$\sigma$ upper limits to the values.

The upper limits to mass-loss rates inferred dynamically are consistent with, but  less stringent than, published mass-loss rates inferred from physical observations. Although $A_2$ is formally significant for 313P/Gibbs, large uncertainties in $A_1$ and $A_3$ degrade the total SNR to $<3$, and therefore only a 5$\sigma$ upper limit for its $\overline{\dot{M}}$ is given in the table. The dynamical estimate for the mass-loss rate of 324P/La Sagra ($36\pm3$ kg s$^{-1}$), however, exceeds  values obtained from physical observations ($\sim$0.2--4 kg s$^{-1}$; Moreno et al. (2011), Hsieh et al. (2012), Jewitt et al. (2016)) by at least an order of magnitude. Notably, while 324P/La Sagra was active, it exhibited the highest ratio of the ejected dust mass to the nucleus mass amongst the active asteroids currently known (Hsieh 2014), suggesting an inherently higher water-ice content. Intriguingly, it is one of the active asteroids identified by Hsieh \& Haghighipour (2016) as a potential captured Jupiter-family comet. This is likely correlated to our finding that 324P/La Sagra has the most significant detection in the non-gravitational acceleration. For (3200) Phaethon, since the detection of its radial non-gravitational acceleration is likely spurious, we only present a 5$\sigma$ upper limit ($< 200$ kg s$^{-1}$)  in Table \ref{tab_phys}. This weak limit is consistent with the perihelion value ($\sim$3 kg s$^{-1}$; Jewitt et al. 2013), as well as the average rate needed to sustain the Geminid stream over its lifetime (Jewitt et al. 2015). In neither case, however, is a firm physical interpretation possible, because it is not known how well the adopted momentum-transfer law represents mass loss that may be highly stochastic in nature.

\subsection{Change in Orbital Elements}

The presence of a non-zero non-gravitational force results in a change of the orbit. Here we proceed to study changes in the semimajor axis, $a$, and eccentricity, $e$, due to the non-gravitational effect, which can be calculated by means of Gauss' form of Lagrange's planetary equations

\begin{align}
\dot{a} & = \frac{P}{\pi} \left[ \mathcal{A}_\mathrm{R} \frac{ e \sin \theta}{\sqrt{1 - e^{2}}} +  \mathcal{A}_\mathrm{T} \frac{a\sqrt{1 - e^{2}}}{r} \right],
\label{eq_adot}
\\
\dot{e} & = \frac{P \sqrt{1 - e^{2}}}{2\pi a} \left[\mathcal{A}_\mathrm{R} \sin \theta + \mathcal{A}_\mathrm{T} \left(\cos \theta + \cos E \right) \right],
\label{eq_edot}
\end{align}

\noindent where $\theta$ is the true anomaly, and $E$ is the eccentric anomaly (Danby 1992). We consider their time-average values by

\begin{align}
\bar{\dot{a}} & \simeq \frac{A_{2} a\sqrt{1 - e^{2}}}{\pi} \int_{0}^{P} \frac{g \left( r \right)}{r} \mathrm{d} t 
\label{eq_dota},
\\
\bar{\dot{e}} & \simeq \frac{A_{2} \sqrt{1 - e^{2}}}{2\pi a} \int_{0}^{P} g \left( r \right) \left[ \cos \theta + \frac{1}{e} \left(1 - \frac{r}{a} \right) \right] \mathrm{d}t
\label{eq_dote},
\end{align}

\noindent Here we have assumed that all of the orbital elements are changing very slowly, such that only $\theta$-dependent functions cannot be taken out of the integral. All the terms containing $\sin \theta$ in the right-hand side of Equations (\ref{eq_adot}) and (\ref{eq_edot}) are eliminated thanks to the orbital symmetry.

By substituting time $t$ with the eccentric anomaly $\theta$ (see Appendix \ref{app_a}), we obtain

\begin{align}
\bar{\dot{a}} & \simeq \frac{P A_{2}}{\pi^{2} a} \int_{0}^{\pi} r g\left( r \right) \mathrm{d} \theta
\label{eq_da},
\\
\bar{\dot{e}} & \simeq \frac{P A_{2}}{2\pi^{2} a^{3}} \int_{0}^{\pi} r^{2} g \left( r \right) \left[ \cos \theta + \frac{1}{e} \left(1 - \frac{r}{a} \right) \right] \mathrm{d}\theta
\label{eq_de},
\end{align}

\noindent Note that Equations (\ref{eq_da}) and (\ref{eq_de}) are only applicable to objects not in strong mean-motion resonances with Jupiter, the most massive planet in the solar system, because the gravitational influence from Jupiter is simply ignored. Indeed, none of the active asteroids are in strong mean-motion resonances with Jupiter. We list the results in Table \ref{tab_phys}. 324P/La Sagra has the most interesting result, with astoundingly large $\bar{\dot{a}}$ and $\bar{\dot{e}}$. The trend indicates that its heliocentric orbit is rapidly becoming smaller and more circular. The timescale to drift $\sim$1 AU, if the non-gravitational effect is persistent, would be $\sim$10$^{5}$ yr. Sustained dynamical evolution on this timescale means that we cannot be sure of the origin of this body, either as a short-period comet trapped from the Kuiper belt or as an icy asteroid from another part of the main-belt.  On the other hand, however, its huge $A_{2}$ suggests a very short active lifetime, limited by the availability of volatiles. Using only physical observations, Jewitt et al.~(2016) reported a lifetime to mass loss of $\sim$10$^{5}$ yr and concluded that, to survive for the expected $\sim$0.4 Gyr collisional lifetime, the body must lie dormant for all but 0.02--0.08\% of the time. In this regard, the inferences from the orbit and from physical observations are concordant.

\subsection{Other Physical Mechanisms}
\label{oth_phys}

We are aware that several mechanisms  other than sublimation account for mass-loss  from some of the active asteroids (Jewitt et al.~2015). While the Yarkovsky effect and the solar radiation pressure force can impart non-gravitational accelerations on an active asteroid in a continuous manner similar to sublimation activity, non-gravitational forces due to rotational instability and impacts obviously cannot be described by the momentum-transfer law in the formalism by Marsden et al. (1973). In particular, mass shedding from rotational instability is believed to be extremely stochastic, as evidenced by distinguishing differences in morphologies between active asteroids possibly experiencing rotational instability (311P/PANSTARRS, 331P/Gibbs, P/2010 A2, and P/2013 R3; Jewitt et al. 2015). We should not expect any detection in non-gravitational effects for these objects, because, first, there is no preference on directions of mass shedding, and second, astrometry from relatively low-resolution observations normally contains larger errors in centroiding optocenters, once there are other fragments apparently close to the primary. Indeed, we have no detections in non-gravitational effects for the active asteroids undergoing suspected rotational instability (see Table \ref{tab_ng}).

The momentum-transfer law by Marsden et al. (1973) also fails for active asteroids suffering from collision-induced mass loss, including (493) Griseldis (Tholen et al. 2015) and (596) Scheila (Ishiguro et al. 2011a,b). The momentum-transfer law for impacts should instead be a Dirac delta function at the time of collision. We investigate changes in the orbital elements for these two active asteroids, considering gravity alone, by comparing the results before and after the impact for each object. No statistically significant detection of orbital change is made. We think that this is in agreement with Ishiguro et al. (2011a) that the impactor ($\sim$10 m) was much smaller than (596) Scheila ($\sim$10$^{2}$ km). For (493) Griseldis, there is unfortunately no size estimate for the impactor.

\subsubsection{Solar Radiation}
The non-gravitational acceleration of a spherical body subjected to solar radiation pressure is given by

\begin{equation}
\left( \mathcal{A}_{\mathrm{R}} \right)_\mathrm{rad} = \frac{3 \left( 1 + \mathrm{A} \right) S_{\odot}}{2 c \rho D r^{2} },
\label{eq_rad}
\end{equation}

\noindent where $c = 3 \times 10^{8}$ m s$^{-1}$ is the speed of light, and $r$ is expressed in AU. We examine the time-average radiation acceleration at mean heliocentric distance $\langle r \rangle = a \sqrt[4]{1 - e^{2}}$ (see Appendix \ref{app_a}) for each active asteroid. If its source is regarded as from  water-ice sublimation, the corresponding radial non-gravitational parameter is then given by $\left( \tilde{A}_{1} \right)_\mathrm{rad} \simeq \left( \overline{\mathcal{A}}_{\mathrm{R}} \right)_\mathrm{rad} / g \left( \langle r \rangle \right)$, where $g \left( r \right)$ remains unchanged from Equation (\ref{eq2}).

We present the results in Table \ref{tab_phys}, where we can see that the observed $A_{1}$ is at least an order of magnitude larger than $\left( \tilde{A}_{1} \right)_\mathrm{rad}$. It therefore suggests that either this effect is too small among the active asteroids, or the uncertainty from the observations is too large to enable such a detection. So far only some near-earth asteroids of $\sim$10 m size have been observed to show measurable acceleration due to solar radiation pressure (e.g.~Micheli et al.~2014). Therefore, we think that the influence of the solar radiation pressure on the (much larger) active asteroids is negligible.

\subsubsection{Yarkovsky Effect}

The other important physical mechanism which can give rise to a non-gravitational acceleration of a sub- or kilometer-sized asteroid is the Yarkovsky effect. Its transverse acceleration is given by

\begin{align}
\nonumber
\left| \left( \mathcal{A}_\mathrm{T} \right)_\mathrm{Y} \right| & = C_\mathrm{Y} \frac{\epsilon \sigma T^3}{c \rho D} \left|  \Delta T \cos \psi \right| \\
 & \le C_\mathrm{Y} \frac{\epsilon \sigma T^3}{c \rho D} \left| \Delta T \right|
\label{eq_ngY}
\end{align}

\noindent where $C_\mathrm{Y}$ is a dimensionless parameter which is related to the object's shape, $\Delta T$ is the temperature difference between the morning and evening hemispheres, and $\psi$ is the obliquity of the object. Thanks to the normalisation to $r = 1$ AU, the relationship $\left(A_{2}\right)_\mathrm{Y} \propto D^{-1}$, where $\left(A_{2}\right)_\mathrm{Y}$ is the transverse non-gravitational parameter due to the Yarkovsky effect, is then roughly satisfied. We therefore use $\left(A_2\right)_\mathrm{Y,Bennu}$, the transverse non-gravitational parameter due to the Yarkovsky effect of asteroid (101955) Bennu, hitherto the most reliable and strongest detection, as a reference to assess expected values for the active asteroids

\begin{equation}
\left| \left(A_2 \right)_\mathrm{Y, exp} \right| = \left| \left(A_2\right)_\mathrm{Y, Bennu} \right| \frac{D_\mathrm{Bennu}}{D}
\label{eq_A2exp},
\end{equation}

\noindent where $\left(A_2\right)_\mathrm{Y, Bennu} = -4.5 \times 10^{-14}$ AU day$^{-2}$, and $D_\mathrm{Bennu} = 0.49$ km is Bennu's diameter (Farnocchia et al. 2013). 

The semimajor-axis drift due to the Yarkovsky effect can be computed by Equation (\ref{eq_da}), with $g(r) = r^{-m}$, where the exact value of $m$ depends upon thermal properties of the asteroid which are, unfortunately,  poorly known. However, the choice of $m$ has little effect in a typical range of $2 < m < 3$ in the computation (Farnocchia et al. 2013), and thus we adopt $m = 2$. Consequently, the expected drift in the semimajor axis can be simplified as

\begin{equation}
\left| \left( \bar{\dot{a} } \right)_\mathrm{Y,exp} \right| \simeq \frac{P \left| \left(A_{2}\right)_\mathrm{Y,Bennu} \right| D_\mathrm{Bennu}}{\pi a^{2} \left( 1 - e^{2}\right) D}
\label{eq_daY_exp}.
\end{equation}

\noindent If the non-gravitational effect of the active asteroid is purely due to the Yarkovsky effect, the criterion $\left| \bar{\dot{a}} \right| \lesssim \left| \left( \bar{\dot{a} } \right)_\mathrm{Y,exp} \right|$ must be satisfied, where $\bar{\dot{a}}$ is listed in Table \ref{tab_phys}. By comparison, we notice that (2201) Oljato, and (3200) Phaethon are the only  two\footnote{Active asteroid (62412) 2000 SY$_{178}$ seemingly satisfies the criterion as well, but it is disqualified by the huge uncertainty in $A_2$ (see Table \ref{tab_ng}). }  potential candidates whose motions might be influenced by the Yarkovsky effect, and we proceed to calculate their $\left(A_{2}\right)_\mathrm{Y}$, by utilising  the same procedures as described in Section \ref{method}.  The results are summarized in Table \ref{tab_Yark}. Unfortunately, neither of the active asteroids show statistically significant detections. We therefore conclude that no Yarkovsky effect is detected amongst the active asteroids. 

It is noteworthy that we failed to reproduce $\left(A_{2}\right)_{\mathrm{Y}}$ of (3200) Phaethon reported by Chernetenko (2010) and Galushina et al. (2015) even though observations after 2015 were discarded as a means to use a similar shorter observing arc. A possible explanation is that they might have assigned too aggressive weights to some of the observations and thus the uncertainty decreases while the nominal $\left(A_2\right)_\mathrm{Y}$ may increase. Instead, our finding of $\left(A_{2}\right)_{\mathrm{Y}}$ of (3200) Phaethon is in good match with D. Farnocchia (2016, private communication).

\section{\uppercase{Summary}}
We examined 18  active asteroids in search of evidence for non-gravitational accelerations caused by anisotropic mass-loss, with the following results:

\begin{enumerate}
\item Three active asteroids (313P/Gibbs, 324P/La Sagra and (3200) Phaethon), exhibit non-gravitational accelerations with at least one component having formal signal-to-noise ratio SNR $> 3$. We are confident in the non-gravitational detections of 313P/Gibbs and, especially, 324P/La Sagra, both kilometer-scale objects with orbital semi-major axes near 3 AU. However, the derived non-gravitational acceleration of (3200) Phaethon, although formally significant, is influenced by systematic uncertainties of measurement, and we do not regard it as real.

\item Upper limits to the mass-loss rates implied by our non-detections of non-gravitational acceleration are less sensitive than, but broadly consistent with, rates inferred independently from physical observations. However, the rate inferred for 324P/La Sagra ($\sim$36 kg s$^{-1}$) is an order of magnitude  larger than values based on physical observations (0.2--4 kg s$^{-1}$).  The reason for this disagreement is not known, but may relate to the poor approximation to impulsive mass loss given by the use of the non-gravitational force law by Marsden et al. (1973).

\item The momentum-transfer law devised by Marsden et al. (1973) assumes sublimation from an isothermal surface and is logically inconsistent with the existence of non-gravitational acceleration (Appendix A). Anisothermal surface temperature distributions are physically more plausible and should replace the law by Marsden et al. (1973). Except in special cases, the law proposed here (Table 5) will give  similar results for the derived non-gravitational parameters.

\item We find no evidence for radiation pressure acceleration or the Yarkovsky effect in our sample.

\end{enumerate}

\acknowledgements
We thank the anonymous referee for  helpful comments and suggestions. This work used the \textit{Find\_Orb} code by Bill Gray, for whose assistance we are extremely grateful. We are indebted to  Aldo Vitagliano, Davide Farnocchia, and Quan-Zhi Ye for insightful discussions. We also thank all observers who submitted astrometric data to the Minor Planet Center, except the ones who submitted really bad astrometry and thus tortured us. This work is funded by a grant from NASA to DJ.

\clearpage

\appendix
\section{\uppercase{The Marsden Momentum Transfer Law}}
\label{marsden}
The momentum-transfer law by Marsden et al. (1973) has been widely used to calculate non-gravitational accelerations of comets. It assumes that sublimation proceeds at a rate appropriate for a uniform, isothermal, spherical nucleus in instantaneous equilibrium with sunlight. However, an isothermal, spherical nucleus would sublimate isotropically, producing no recoil force. Therefore, the law by Marsden et al. (1973) is logically inconsistent with the presence of non-gravitational acceleration. We briefly examine the significance of this inconsistency.

\begin{figure}
\epsscale{1.0}
\begin{center}
\plotone{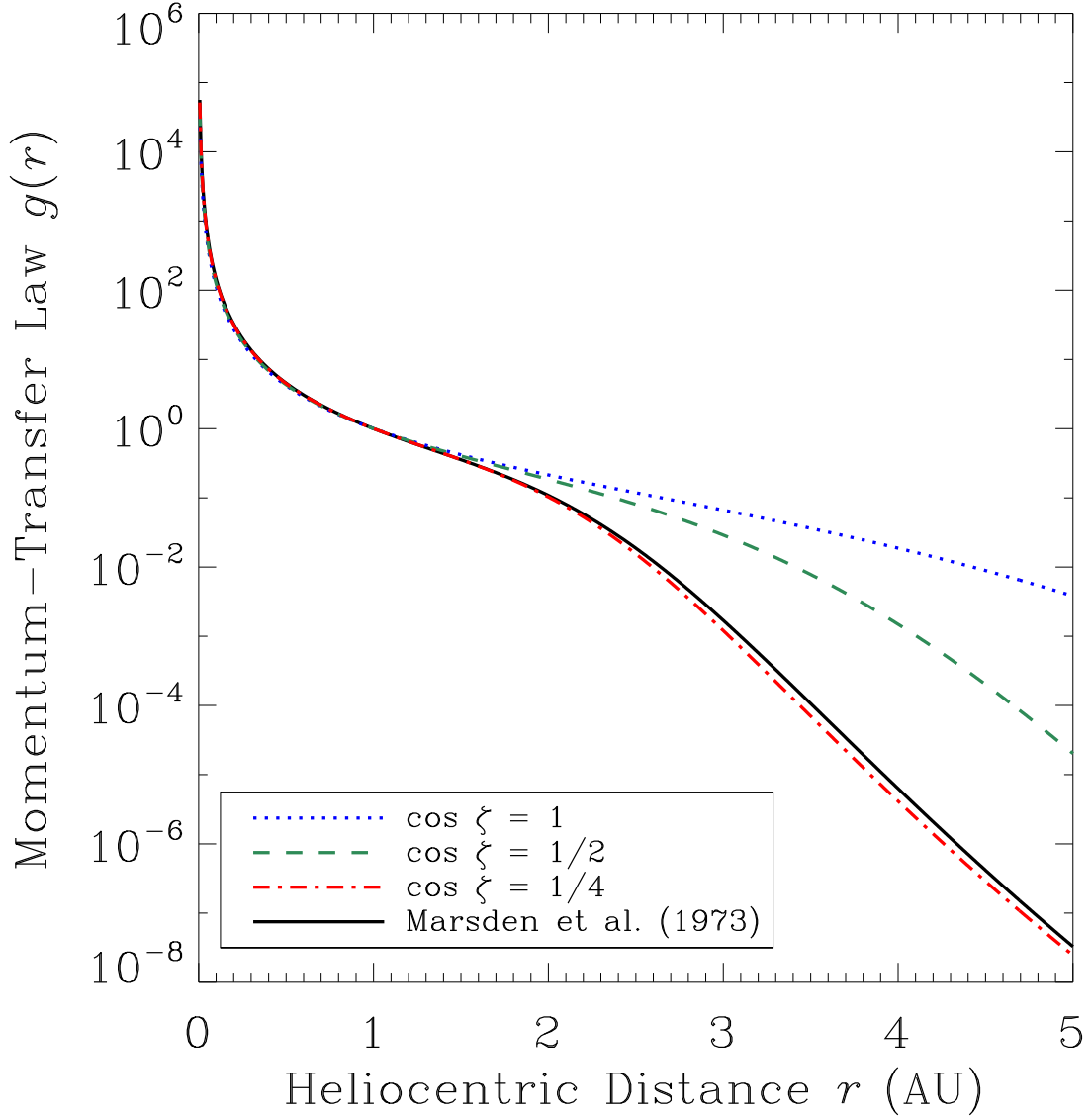}
\caption{
Comparison of our best fits in the formalism by Equation (\ref{eq2}) for three different sublimation scenarios, i.e., $\cos \zeta = 1/4$ (isothermal sublimation), 1/2, and 1 (subsolar), and the best fit by Marsden et al. (1973). The actual normalised water-ice sublimation functions are indistinguishable from our best fits correspondingly, were they plotted in the figure, and therefore are omitted. Different fits are discriminated by line styles. 
\label{fig_glaw}
} 
\end{center} 
\end{figure}

As limiting cases, we compare in Figure (\ref{fig_glaw}) the model by Marsden et al. (1973) (solid black line) with three different solutions to Equation (\ref{eq_sub}). Our approximation to isothermal sublimation (labeled $\cos \zeta = 1/4$ and shown by a red dash-dot line in the figure) essentially reproduces that by Marsden et al. (1973). Models in which sunlight heats only the day-side of the nucleus ($\cos \zeta = 1/2$, dashed green line) and in which heat is deposited only at the sub-solar point ($\cos \zeta = 1$, dotted blue line) both show substantially higher specific sublimation rates at $r \gtrsim 2.5$ AU as a result of the higher average temperatures. The revised non-gravitational parameters for these models are listed in Table \ref{tab_glaw}. 

To test the effect of the differences shown in Figure (\ref{fig_glaw}), we computed new orbits of selected short-period and Halley-type comets with nonzero non-gravitational effects\footnote{This was checked through the JPL Small-Body Database Search Engine. Only comets with $>$$10\sigma$ detections on non-gravitational effects were selected.} using astrometric data from the MPC Database Search with the parameters in Table \ref{tab_glaw}. We found that, even when using the two most extreme scenarios (namely, the isothermal ($\cos \zeta  = 1/4$) and subsolar ($\cos \zeta = 1$) models), the derived orbital solutions and time-average non-gravitational accelerations are unchanged, within the uncertainties. Specifically, the RMS of best fits computed using the different momentum transfer laws of Table \ref{tab_glaw} are basically the same. Physically, this is because the differences between the sublimation curves in Figure (\ref{fig_glaw}) are significant only at $r \gtrsim$ 2.5 AU, where the momentum flux driven by water-ice sublimation is already very low. Nevertheless, our suggestion is for future work to use the best-fit parameters given in Table \ref{tab_glaw} for $\cos \zeta = 1/2$. This case is physically the most plausible, since cometary nuclei are observed to sublimate primarily from the dayside (Keller et al. 2004), and it is also logically consistent with a net force acting on the nucleus. 

Of course in reality, non-gravitational effects due to mass-loss activity are strongly dependent on, for instance, the shape, topography,  spin, and thermal properties of individual nuclei, as well as the distribution of volatiles. It is impractical to devise a model which can universally satisfy all the cases of such complexity. Besides, little is known about the nuclei of the majority of comets. Therefore, adopting the aforementioned simplistic model is still appropriate and necessary for most cases.

\section{\uppercase{Derivation of Time-Average Values}}
\label{app_a}
Let us consider a continuous function of time $t$ which is symmetric about axes of a body's elliptical orbit, denoted as $f \left(t \right)$. The elliptical orbit has semimajor axis $a$ and eccentricity $e$. Now the task is to find its time-average value

\begin{equation}
\bar{f} = \frac{1}{P} \int_{0}^{P} f \left( t \right) \mathrm{d} t ,
\label{eq_A1}
\end{equation}

\noindent where $P$ is the orbital period. Because $f \left( t \right)$ is symmetric about the axes of the ellipse, i.e., $f \left(P - t \right) = f\left( t \right)$, Equation (\ref{eq_A1}) is therefore equivalent to

\begin{equation}
\bar{f} = \frac{2}{P} \int_{0}^{\frac{P}{2}} f \left( t \right) \mathrm{d} t .
\label{eq_A2}
\end{equation}

It is often the case where $f$ is explicitly a function of true anomaly $\theta$, i.e., $f = f \left(\theta \right)$, and henceforth we need to find a way which connects $\theta$ and $t$. From orbital mechanics we know the following relationships:

\begin{align}
\label{eq_A3}
t - t_{0} & =  \frac{P}{2\pi} M ,\\ 
\label{eq_A4}
M &= E - \sin E , \\ 
\label{eq_A5}
E &= \arccos \left( \frac{e + \cos \theta}{1 + e \cos \theta} \right),
\end{align}

\noindent where $M$ is the mean anomaly, and $E$ is the eccentric anomaly. Differentiating both sides from Equation (\ref{eq_A3}) to (\ref{eq_A5}) yields

\begin{align}
\label{deq_A3}
\mathrm{d} t & =  \frac{P}{2\pi} \mathrm{d} M ,\\ 
\label{deq_A4}
\mathrm{d} M &= \left( 1 - \cos E \right) \mathrm{d} E , \\ 
\label{deq_A5}
\mathrm{d} E &= \frac{\sqrt{1 - e^{2}}}{1 + e \cos \theta} \mathrm{d} \theta .
\end{align}

We then apply the chain rule to Equation (\ref{eq_A2}) and obtain

\begin{align}
\nonumber
\bar{f} & = \frac{2}{P} \int_{0}^{\pi} \mathrm{d} \theta \frac{\mathrm{d} E}{\mathrm{d} \theta} \frac{\mathrm{d}M}{\mathrm{d}E} \frac{\mathrm{d}t}{\mathrm{d}M} f \\
 & = \frac{\left(1 - e^{2} \right)^{3/2}}{\pi} \int_{0}^{\pi} \mathrm{d} \theta \frac{f \left( \theta \right)}{\left(1 + e \cos \theta\right)^{2}} .
 \label{eq_A9}
\end{align}

 Under polar coordinates with one of the foci at the origin, which represents the Sun, and the other focus on the negative $x$-axis, the elliptical orbit is expressed by

\begin{equation}
r = \frac{a \left( 1 - e^{2} \right)}{1 + e \cos \theta}.
\label{eq_A10}
\end{equation}

\noindent Combining Equations (\ref{eq_A9}) with (\ref{eq_A10}), we derive

\begin{equation}
\bar{f} = \frac{1}{\pi a^{2} \sqrt{1 - e^{2}}} \int_{0}^{\pi} \mathrm{d} \theta f\left( \theta \right) r^{2}.
 \label{eq_A11}
\end{equation}

In this study we need mean temperatures of the active asteroids, whose orbits are approximately elliptic, by ignoring perturbations from other bodies and non-gravitational effects. In accordance with Equation (\ref{eq_sub}), we have $f = r^{-2}$ in this scenario. Immediately, we obtain

\begin{equation}
\overline{\left( \frac{1}{r^{2}}\right)} = \frac{1}{a^{2} \sqrt{1 - e^{2}}}.
\end{equation}

\noindent The equivalent mean heliocentric distance under this definition is thereby $\langle r \rangle = a \sqrt[4]{1 - e^{2}}$. Interestingly, the time-average heliocentric distance is $\bar{r} = a \left(1 + e^{2} / 2 \right)$, given by Equation (\ref{eq_A9}) with $f = r$.









\begin{deluxetable}{lcccccccccc}
\tabletypesize{\scriptsize}
\rotate
\tablecaption{Non-Gravitational Parameters of Active Asteroids \label{tab_ng}}
\tablewidth{0pt}
\tablehead{
\colhead{Object}    & \colhead{$A_{1}$}  &  \colhead{SNR($A_1$)} & \colhead{$A_{2}$}   &  \colhead{SNR($A_2$)} & \colhead{$A_{3}$}  & \colhead{SNR($A_3$)}
& \colhead{Data arc} & \colhead{\# obs\tablenotemark{\dagger}} & \colhead{\# opp\tablenotemark{\ddagger}} & \colhead{RMS}\\
  & \colhead{(AU day$^{-2}$)} & & \colhead{(AU day$^{-2}$)} & & \colhead{(AU day$^{-2}$)}  & & & & & \colhead{(\arcsec)}

} 

\startdata
107P & $-1.15 \times 10^{-11}$ & 2.03 & $-3.56 \times 10^{-14}$ & 2.58 & $+1.64 \times 10^{-11}$ & 1.97 & 1949--2016 & 909 (17) & 18 & 0.57 \\ 

133P & $+5.09 \times 10^{-10}$ & 2.62 & $+3.63 \times 10^{-12}$ & 0.33 & $-1.14 \times 10^{-10}$ & 0.33 & 1979--2016 & 716 (13) & 18 & 0.50 \\ 

176P & $-4.83 \times 10^{-10}$ & 2.64 & $-1.04 \times 10^{-11}$ & 0.42 & $-9.12 \times 10^{-11}$ & 0.18 & 1999--2016 & 568 (2) & 14 & 0.48 \\ 

238P
& $-4.18 \times 10^{-8}$ & 1.13 & $-3.40 \times 10^{-8}$ & 2.13 & $+6.12 \times 10^{-12}$ & $< 1$\% & 2005--2011 & 141 (0) & 4 & 0.59 \\ 

259P
& $-2.88 \times 10^{-8}$ & 2.97 & $+5.17 \times 10^{-9}$ & 0.70 & $+1.10 \times 10^{-8}$ & 2.61 & 2008--2012 & 40 (6) & 4 & 0.73 \\ 

288P
& $ -1.26 \times 10^{-10}$ & 0.20 & $ +4.69 \times 10^{-12}$ & 0.09 & $ -5.31 \times 10^{-10}$ & 1.38 & 2000--2015 & 160 (0) & 9 & 0.52 \\ 

311P
& $ +2.28 \times 10^{-9}$ & 1.85 & $ +3.12 \times 10^{-11}$ & 2.23 & $ -6.36 \times 10^{-10}$ & 1.10 & 2005--2015 & 158 (3) & 5 & 0.45 \\ 

313P
& $ +3.27 \times 10^{-8}$ & 1.75 & $ +2.13 \times 10^{-8}$ & 4.45 & $ -4.82 \times 10^{-9}$ & 1.83 & 2003--2014 & 94 (3) & 3 & 0.63 \\ 

324P & $ -2.96 \times 10^{-7}$ & 10.46 & $ -1.47 \times 10^{-7}$ & 10.50 & $ -3.75 \times 10^{-8}$ & 7.41 & 2010--2015 & 421 (2) & 4 & 0.48 \\ 

331P & $ -1.09 \times 10^{-7}$ & 2.24 & $ +5.16 \times 10^{-10}$ & 0.87 & $ +6.58 \times 10^{-9}$ & 0.96 & 2004--2015 & 148 (10) & 6 & 0.86 \\ 

493 & $ +6.71 \times 10^{-11}$ & 0.73 & $ -2.47 \times 10^{-12}$ & 1.80 & $ +1.74 \times 10^{-12}$ & 0.01 & 1902--2016 & 1388 (29) & 27 & 0.51 \\ 

596 & $ +7.53 \times 10^{-12}$ & 0.22 & $ -1.16 \times 10^{-12}$ & 1.75 & $ -1.85 \times 10^{-10}$ & 2.14 & 1908--2016 & 3418 (71) & 41 & 0.40 \\ 

2201 & $ +4.67 \times 10^{-13}$ & 0.15 & $ +2.95 \times 10^{-14}$ & 2.29 & $ -3.36 \times 10^{-12}$ & 0.32 & 1931--2015 & 823 (23) & 25 & 0.51 \\ 

3200 & $ +6.97 \times 10^{-12}$ & 3.40 & $ -1.44 \times 10^{-15}$ & 0.92 & $ +8.88 \times 10^{-13}$ & 0.59 & 1983--2016 & 3161 (60) & 30 & 0.46 \\ 

62412 & $ +5.20 \times 10^{-10}$ & 0.83 & $ -1.53 \times 10^{-14}$ & $< 1$\% & $ +1.02 \times 10^{-9}$ & 1.08 & 1999--2016 & 737 (2) & 13 & 0.54 \\ 

P/2010 A2
& $ -1.76 \times 10^{-7}$ & 1.56 & $ +7.97 \times 10^{-8}$ & 2.21 & $ -1.10 \times 10^{-7}$ & 1.34 & 2010--2012 & 127 (95) & 2 & 1.23 \\ 

P/2012 T1
& $ -6.52 \times 10^{-6}$ & 1.42 & $ -1.06 \times 10^{-6}$ & 1.58 & $ +2.22 \times 10^{-7}$ & 1.27 & 2012--2013 & 165 (1) & 1 & 0.45 \\ 

P/2013 R3
& $ +1.65 \times 10^{-6}$ & 1.04 & $ +6.80 \times 10^{-7}$ & 1.40 & $ -5.23 \times 10^{-8}$ & 2.19 & 2013--2014 & 316 (5) & 1 & 0.63 

\enddata

\tablenotetext{\dagger}{Total number of observations of all types (optical and radar) used in fit. Number of discarded data bracketed. }
\tablenotetext{\ddagger}{Number of observed oppositions}

\tablecomments{
The non-gravitational parameters are calculated based on the isothermal water-ice sublimation model devised by Marsden et al. (1973). SNR($A_j$) ($j=1,2,3$) is the ratio of $\left| A_j \right|$ over its 1$\sigma$ uncertainty. All of the astrometric observations were retrieved on 2016 July 14--15.
}

\end{deluxetable}

\clearpage

\begin{deluxetable}{lcccccccccc}
\tabletypesize{\scriptsize}
\rotate
\tablecaption{Non-Gravitational Parameters of Some Moderate-Size Asteroids \label{tab_ast10km}}
\tablewidth{0pt}
\tablehead{
\colhead{Object}    & \colhead{$A_{1}$}  &  \colhead{SNR($A_1$)} & \colhead{$A_{2}$}   &  \colhead{SNR($A_2$)} & \colhead{$A_{3}$}  & \colhead{SNR($A_3$)}
& \colhead{Data arc} & \colhead{\# obs\tablenotemark{\dagger}} & \colhead{\# opp\tablenotemark{\ddagger}} & \colhead{RMS}\\
  & \colhead{(AU day$^{-2}$)} & & \colhead{(AU day$^{-2}$)} & & \colhead{(AU day$^{-2}$)}  & & & & & \colhead{(\arcsec)} 
}

\startdata

3818 & $ -2.73 \times 10^{-11}$ & 1.65 & $ +2.31 \times 10^{-13}$ & 0.62 & $ +3.19 \times 10^{-11}$ & 1.16 & 1979--2015 & 1166 (16) & 20 & 0.49 \\

7916 & $ -3.29 \times 10^{-12}$ & 0.23 & $ +1.72 \times 10^{-13}$ & 0.61 & $ +2.56 \times 10^{-11}$ & 1.03 & 1978--2015 & 1080 (5) & 18 & 0.53 \\

9344 & $ +1.86 \times 10^{-11}$ & 0.58 & $ +1.93 \times 10^{-12}$ & 2.59 & $ +1.12 \times 10^{-10}$ & 2.45 & 1991--2016 & 1222 (6) & 16 & 0.54 \\

11313 & $ +6.59 \times 10^{-12}$ & 0.09 & $ -1.48 \times 10^{-12}$ & 1.69 & $ +1.67 \times 10^{-10}$ & 1.90 & 1976--2016 & 1219 (3) & 18 & 0.52 \\

13426 & $ -7.85 \times 10^{-12}$ & 0.45 & $ -2.37 \times 10^{-13}$ & 0.86 & $ +1.97 \times 10^{-11}$ & 0.71 & 1975--2015 & 792 (2) & 14 & 0.54 \\

16392 & $ -9.26 \times 10^{-11}$ & 1.84 & $ -2.37 \times 10^{-13}$ & 0.13 & $ +5.85 \times 10^{-11}$ & 0.57 & 1977--2016 & 1085 (2) & 19 & 0.50 \\

18333 & $ +2.52 \times 10^{-11}$ & 1.02 & $ -1.20 \times 10^{-13}$ & 0.09 & $ +3.83 \times 10^{-11}$ & 0.71 & 1987--2016 & 1100 (4) & 16 & 0.54 \\

20099 & $ -6.45 \times 10^{-12}$ & 0.05 & $ -5.11 \times 10^{-12}$ & 0.51 & $ +2.97 \times 10^{-11}$ & 0.14 & 1991--2015 & 852 (1) & 17 & 0.49 \\

20293 & $ +3.16 \times 10^{-11}$ & 2.09 & $ -9.05 \times 10^{-13}$ & 1.66 & $ +6.33 \times 10^{-11}$ & 1.94 & 1980--2016 & 1099 (5) & 15 & 0.52 \\

23059 & $ +6.32 \times 10^{-13}$ & 0.05 & $ -3.62 \times 10^{-13}$ & 0.76 & $ +1.97 \times 10^{-11}$ & 0.80 & 1991--2016 & 1240 (1) & 15 & 0.47 \\

25343 & $ -1.36 \times 10^{-11}$ & 0.50 & $ -1.66 \times 10^{-12}$ & 2.54 & $ +7.86 \times 10^{-11}$ & 1.95 & 1992--2015 & 866 (6) & 16 & 0.56 \\

26662 & $ +4.64 \times 10^{-11}$ & 1.08 & $ -4.82 \times 10^{-13}$ & 0.62 & $ +7.31 \times 10^{-11}$ & 2.09 & 1974--2015 & 636 (1) & 17 & 0.56 \\

\enddata

\tablenotetext{\dagger}{Total number of observations of all types (optical and radar) used in fit. Number of discarded data bracketed. }
\tablenotetext{\ddagger}{Number of observed oppositions}

\tablecomments{
All of the asteroids have diameters $\sim$10 km. The non-gravitational parameters are calculated based on the isothermal water-ice sublimation model devised by Marsden et al. (1973). All of the astrometric observations were retrieved on 2016 July 14--15.
}

\end{deluxetable}

\begin{deluxetable}{lcccrrcc}
\tabletypesize{\footnotesize}
\tablecaption{Physical and Derived Properties \label{tab_phys}}
\tablewidth{0pt}
\tablehead{
\colhead{Object}    & \colhead{$D$\tablenotemark{(1)}}  &  
\colhead{$\mathrm{A}$\tablenotemark{(2)}} & \colhead{$-\overline{\dot{M}}$\tablenotemark{(3)}} & 
\colhead{$\left( \overline{\mathcal{A}}_\mathrm{R} \right)_\mathrm{rad}$\tablenotemark{(4)}} & 
\colhead{$\left( \widetilde{A}_{1} \right)_\mathrm{rad}$\tablenotemark{(5)}}
& \colhead{$\bar{\dot{a}}$\tablenotemark{(6)}} 
& \colhead{$\bar{\dot{e}}$\tablenotemark{(7)}}  \\
& \colhead{(km)} & & \colhead{(kg s$^{-1}$)} 
& \colhead{(AU day$^{-2}$)} & \colhead{(AU day$^{-2}$)} 
& \colhead{(AU yr$^{-1}$)} & \colhead{(yr$^{-1}$)} \\

} 

\startdata
107P & 3.5 & 0.02 & $< 5$  & 
$1.82 \times 10^{-14}$ & $5.06\times10^{-13}$ 
& $-1.9\times10^{-9}$ & $-2.7\times10^{-10}$ \\

133P & 3.8   & 0.02  & $< 4$  & 
$9.26 \times 10^{-15}$ & $1.14\times10^{-11}$ 
& $+3.2\times10^{-9}$ & $+6.5\times10^{-10}$ \\

176P & 4.0   & 0.02  & $< 5$  &  
$8.68 \times 10^{-15}$ & $1.24\times10^{-11}$ & 
$-1.2\times10^{-8}$ & $-2.5\times10^{-9}$ \\

238P & 0.8  & 0.02 & $< 13$  &  
$4.48 \times 10^{-14}$ & $4.76\times10^{-11}$ 
& $-9.9\times10^{-5}$ & $-2.0\times10^{-5}$ \\

259P & 0.6   & 0.02  & $< 32$  & 
$8.25 \times 10^{-14}$ & $8.35\times10^{-12}$ 
& $+6.7\times10^{-5}$ & $+1.3\times10^{-5}$ \\

288P & 3  & 0.02 & $< 8 $
& $1.26 \times 10^{-14}$ & $8.14\times10^{-12}$ 
& $+9.5\times10^{-9}$ & $+2.0\times10^{-9}$ \\

311P & $<0.5$  &  0.11 & $< 1 $ 
& $> 1.59 \times 10^{-13}$ & $>2.56\times10^{-12}$ 
& $+3.1\times10^{-7}$ & $+4.1\times10^{-8}$ \\

313P & 1.0  & 0.02 & $< 12 $
& $3.59 \times 10^{-14}$ & $3.76\times10^{-11}$ 
& $+5.4\times10^{-5}$ & $+1.1\times10^{-5}$ \\

324P & 1.1  & 0.02  & $36\pm3$  
& $3.31 \times 10^{-14}$ & $2.95\times10^{-11}$ 
& $-1.4\times10^{-4}$ & $-2.8\times10^{-5}$ \\

331P & 1.8  & 0.02 & $\lesssim 77$  
& $2.13 \times 10^{-14}$ & $1.28\times10^{-11}$ 
& $+2.1\times10^{-7}$ & $+2.0\times10^{-8}$ \\

493 & 46.4   & 0.02 &  $\lesssim10^{3}$  
& $7.81 \times 10^{-16}$ & $7.84\times10^{-13}$ 
& $-2.7\times10^{-9}$ & $-5.5\times10^{-10}$ \\

596  & 113.3   & 0.01 & $\lesssim10^{5}$  
& $3.60 \times 10^{-16}$ & $1.30\times10^{-13}$ 
& $-2.2\times10^{-9}$ & $-4.5\times10^{-10}$ \\

2201 & 1.8    & 0.17  & $< 2  $
& $6.68 \times 10^{-14}$ & $3.81\times10^{-13}$ 
& $+2.8\times10^{-9}$ & $+3.8\times10^{-10}$ \\

3200 & 5.1    & 0.04  & $< 200 $
& $9.36 \times 10^{-14}$ & $6.66\times10^{-14}$ 
& $-9.4\times10^{-10}$ & $-2.6\times10^{-10}$ \\

62412 & 7.8   & 0.03  & $< 70$  
& $4.51 \times 10^{-15}$ & $5.87\times10^{-12}$ 
& $-4.6\times10^{-12}$ & $-7.3\times10^{-13}$ \\

P/2010 A2 & 0.12  & 0.04  & $< 1  $
& $5.65 \times 10^{-13}$ & $1.29\times10^{-11}$ 
& $+6.5\times10^{-4}$ & $+9.8\times10^{-5}$ \\

P/2012 T1 & 2.4  & 0.02 & $\lesssim10^{4}$  
& $1.49 \times 10^{-14}$ & $1.57\times10^{-11}$ 
& $-2.5\times10^{-3}$ & $-5.2\times10^{-4}$ \\

P/2013 R3 & $<0.4$  & 0.02  & $< 141$  
& $>9.77 \times 10^{-14}$ & $>5.09\times10^{-11}$ 
& $+3.3 \times 10^{-3}$ & $+6.7 \times 10^{-4}$ \\

\enddata

\tablenotetext{(1)}{Diameter}
\tablenotetext{(2)}{Bond albedo}
\tablenotetext{(3)}{Time-average mass-loss rate estimated from Equation (\ref{eq_mloss2})}
\tablenotetext{(4)}{Computed non-gravitational acceleration due to the solar radiation force}
\tablenotetext{(5)}{Radial non-gravitational parameter due to the solar radiation force but computed with the momentum-transfer law by Marsden et al. (1973)}
\tablenotetext{(6)}{Time-average drift in semimajor axis}
\tablenotetext{(7)}{Time-average drift in eccentricity}

\tablecomments{
The significance levels of an orbital drift in $a$ and $e$ are predominantly determined by the ones of the non-gravitational parameters, which are the most uncertain parameters compared to the rest orbital elements. See Equations (\ref{eq_da}) and (\ref{eq_de}). Therefore, the SNRs of $\bar{\dot{a}}$ and $\bar{\dot{e}}$ are both given by SNR($A_2$), listed in Table \ref{tab_ng}.
}

\end{deluxetable}

\begin{deluxetable}{lcrcccc}
\tablecaption{Transverse Non-Gravitational Parameters Due to the Yarkovsky Effect \label{tab_Yark}}
\tablewidth{0pt}
\tablehead{
\colhead{Object}  & \colhead{$\left|\left(A_{2}\right)_\mathrm{Y, exp}\right|$\tablenotemark{\dagger}} &  \colhead{$\left(A_{2}\right)_\mathrm{Y}$\tablenotemark{\ddagger}}  
& \colhead{Data arc} & \colhead{\# obs\tablenotemark{\ast}} & \colhead{\# opp\tablenotemark{\star}} & \colhead{RMS}\\
  & \colhead{(AU day$^{-2}$)} & \colhead{(AU day$^{-2}$)} 
  & & & & \colhead{(\arcsec)}
} 

\startdata
2201  & $1.2 \times 10^{-14}$ & $\left(+2.89 \pm 1.28 \right) \times 10^{-14}$ & 1931--2015 & 824 (22) & 25 & 0.51 \\ 

3200 & $4.4 \times 10^{-15}$ & $\left(-1.39 \pm 1.56 \right) \times 10^{-15}$ & 1983--2016 & 3161 (60) & 30 & 0.46 \\ 


\enddata

\tablenotetext{\dagger}{Value of expected transverse non-gravitational parameter due to the Yarkovsky effect estimated from the one of (101955) Bennu through Equation (\ref{eq_A2exp}).}
\tablenotetext{\ddagger}{Transverse non-gravitational parameter due to the Yarkovsky effect computed from orbit determination.}
\tablenotetext{\ast}{Total number of observations of all types (optical and radar) used in fit. Number of discarded data bracketed. }
\tablenotetext{\star}{Number of observed oppositions. }

\tablecomments{
The same technique as used for obtaining the non-gravitational parameters in Table \ref{tab_ng} is applied, with the modified momentum-transfer law $g(r) = r^{-2}$.
}

\end{deluxetable}


\begin{deluxetable}{ccccc}

\tablecaption{Parameters in the Momentum-Transfer Law \label{tab_glaw}}
\tablewidth{0pt}
\tablehead{
\colhead{Parameter} & \colhead{$\cos \zeta = 1/4$} & \colhead{$\cos \zeta = 1/2$} & \colhead{$\cos \zeta = 1$} & \colhead{Unit} \\
 & \colhead{(Isothermal)} & \colhead{(Hemispherical)} & \colhead{(Subsolar)} &
}

\startdata
$\alpha$ & 0.1258295 & 0.0337694 & 0.0003321 & -- \\
$m$ & 2.13294 & 2.08782 &  2.04680 & -- \\
$n$ & 5.30728 &  4.04051 &  3.06682 & -- \\
$k$ & 4.19724 & 11.4543 &  2752.35& -- \\
$r_0$ & 2.67110 & 5.10588 &  50.4755 & AU \\

\enddata

\tablecomments{
Each least-squares fit was performed for heliocentric distance $r \le 5$ AU, beyond which the contribution from the water-ice sublimation is negligible. See Figure \ref{fig_glaw} for comparison.
}

\end{deluxetable}

\end{document}